\documentclass[12pt,reqno]{article}
\usepackage{amsmath,hyperref,amssymb}
\usepackage{subfigure,graphicx,url}

\allowdisplaybreaks \numberwithin{equation}{section}

\DeclareSymbolFont{AMSa}{U}{msa}{m}{n}
\DeclareSymbolFont{AMSb}{U}{msb}{m}{n}
\DeclareMathSymbol{\fieldR}{\mathalpha}{AMSb}{"52}

\begin{document}

\begin{flushright} \small
ITP--UU--10/48 \\ SPIN--10/41
\end{flushright}
\bigskip
\begin{center}
 {\large\bfseries Static supersymmetric black holes in AdS$_4$ with spherical symmetry}\\[5mm]
Kiril Hristov$^{*,\dag}$, Stefan Vandoren$^*$ \\[3mm]
 {\small\slshape
 * Institute for Theoretical Physics \emph{and} Spinoza Institute, \\
 Utrecht University, 3508 TD Utrecht, The Netherlands \\
\medskip
 \dag Faculty of Physics, Sofia University, Sofia 1164, Bulgaria\\
\medskip
 {\upshape\ttfamily K.P.Hristov, S.J.G.Vandoren@uu.nl}\\[3mm]}
\end{center}
\vspace{5mm} \hrule\bigskip \centerline{\bfseries Abstract}
\medskip

We elaborate further on the static supersymmetric AdS$_4$ black
holes found in \cite{klemm-adsBH}, investigating thoroughly the
BPS constraints for spherical symmetry in $N=2$ gauged
supergravity in the presence of Fayet-Iliopoulos terms. We find
Killing spinors that preserve two of the original
eight supercharges and investigate the conditions for genuine
black holes free of naked singularities. The existence of a
horizon is intimately related with the requirement that the
scalars are not constant, but given in terms of harmonic functions
in analogy to the attractor flow in ungauged supergravity. The
black hole charges depend on the choice of the electromagnetic gauging, with only
magnetic charges for purely electric gaugings.
Finally we show how these black holes can be
embedded in $N=8$ supergravity and thus in M-theory.

\bigskip
\hrule\bigskip

\section{Introduction}\label{intro}

The study of black holes in supergravity and string theory has
been of general interest for many years. Research topics range
from fundamental aspects of quantum gravity and microscopic state
counting in string theory, to applications of black hole
thermodynamics in strongly coupled field theories via the AdS/CFT
correspondence. Many properties of black holes depend on the
asymptotic spacetime they live in, which can be flat, de Sitter,
or anti-de Sitter (AdS). Most studies focus on asymptotically flat
or AdS spacetimes, and in this work we focus on the latter.

In this paper we analyze a class of static supersymmetric (BPS),
asymptotically AdS$_4$ black holes with a spherical horizon in
gauged $N=2$ supergravity. Static BPS solutions with other horizon
topologies, as well as stationary rotating solutions, are known to
exist for long time in such theories
\cite{Caldarelli:1998hg,Kostelecky:1995ei}\footnote{Note that,
unlike the case for asymptotically flat static black holes, the
topology of the horizon of AdS$_4$ black holes is not unique. The
horizon can be a Riemann surface of any genus as explained in
\cite{Caldarelli:1998hg}.}. However, until recently static BPS
black holes with spherical horizons were thought not to exist, at
least not for the choices of gauging and Killing spinor
ans\"{a}tze studied in  e.g. \cite{sabra-ads}. While this is the
case in minimally gauged supergravity with a bare cosmological
constant \cite{Romans:1991nq}, the first example of proper static
BPS solution in the presence of vector multiplets and a scalar
potential was derived in \cite{klemm-adsBH}, building on earlier
work \cite{klemm}.

Just like in \cite{klemm-adsBH}, we concentrate on gaugings with
Fayet-Iliopoulos (FI) terms. We do not consider hypermultiplets,
but in certain cases the hypermultiplet gaugings allow for
truncations to the models we consider here \cite{our}. As we will
explain, the FI terms determine the electric charges of the
gravitini and are subject to a Dirac quantization in the presence
of any magnetic charges. The black holes we study in this paper
are magnetically charged, and have an entropy that depends on both
magnetic charges and FI terms. The fact that they are quantized
will therefore be important for the microscopic state counting.

The complete set of BPS conditions were written down in
\cite{klemm-adsBH}, with no constraint on the topology of the
horizon and no assumption on the form of the Killing spinors.
While this covers the most general case, the equations are
somewhat cumbersome and difficult to analyze unless one specifies
to detailed examples. Here, we aim to understand better the case
of spherical horizons only, for which the BPS conditions simplify
once we restrict to a particular class of Killing spinors. In this
way, one recovers attractor-like equations that are similar to the
ones describing asymptotically flat black holes in ungauged
supergravity
\cite{Ferrara:1995ih,Strominger:1996kf,Ferrara:1996dd}. We also
extend the analysis beyond the standard electrically gauged $N=2$
supergravity, by allowing magnetic gaugings. In such models, we
can describe more general black hole solutions that have both
electric and magnetic charges on equal footing.

As an illustration, we consider the case of one vectormultiplet.
This example was also studied in \cite{klemm-adsBH}, where a
spherically symmetric black hole with no naked singularity was
found. We discuss further the properties of this black hole, such
as the entropy formula and the attractor mechanism. Furthermore,
we also comment on the mass of the black hole and describe the
embedding into eleven-dimensional supergravity.

The plan of the paper is as follows. First, in section
\ref{overview} we discuss the known static AdS black holes in four
dimensions and explain how the solutions described in this paper
fit in the general picture. In section \ref{sect:2} we briefly
outline some details about gauged $N=2$ supergravity and explain
our notations. In section \ref{sect:3} we specify in full detail
our assumptions for spacetime and gauge fields and make a
particular ansatz for the Killing spinors in order to simplify the
BPS conditions. In section \ref{sect:4} we show how to solve the
equations for the metric and scalar fields in terms of harmonic
functions. We then proceed in section \ref{electr-magn general
solution} to explain how the embedding tensor formalism
\cite{embedding} restores electromagnetic duality and propose a
more general solution in an arbitrary electromagnetic frame. In
section \ref{sect:explcit_examples} we give some explicit examples
of prepotentials leading to black hole solutions and give more
details about the physical properties of these black holes and the
attractor flow. In the last part of the paper, we show in section
\ref{sect:M-theory reduction} how one can embed these new black
holes in $D =4$ $N=8$ supergravity and in M-theory, thus
suggesting a way to study their microscopic origin. In section
\ref{sect:Thermo} we comment on the mass of the black holes and
their behavior in the large charge limit, which shows some quite
unusual and puzzling behavior. We conclude with some further
remarks and suggestions for future study. Some details about our
gamma matrix conventions are left for the Appendix.

Note added: Just before this paper was submitted, we received reference \cite{Dall'Agata} that has some overlap
with our results in the sections discussing the dyonic solutions with electromagnetic gauging and the attractor
mechanism.

\section{Static AdS black holes}\label{overview}

We focus on static spherically symmetric spacetimes with metrics of the form
(the signature is $(+,-,-,-)$ in our conventions)
\begin{equation}
{\rm d} s^2 = U^2(r)\, {\rm d}t^2 - U^{-2}(r)\, {\rm d}r^2 -
h^2(r)\, ({\rm d} \theta^2 + \sin^2 \theta {\rm d} \varphi^2)\ ,
\end{equation}
for some functions $U(r)$ and $h(r)$ to be determined from the BPS
conditions and/or the equations of motion.

For Minkowski spacetime, we have $U=1$ and $h=r$, and for
four-dimensional anti-de Sitter spacetime, one has
\begin{equation}
AdS_4: \qquad U^2(r)=1+g^2 r^2\ ,\qquad h(r)= r\ ,
\end{equation}
where $g$ is related to the
cosmological constant of AdS$_4$ through the scalar curvature relation
$R=-12 g^2$. So, in the standard conventions the cosmological
constant is $\Lambda = -3 g^2$. For the Reisnner-Nordstr\"om
black hole solution in AdS$_4$ (RN-AdS), with mass $M$ and
electric and magnetic charges $Q$ and $P$, we have
\begin{equation}\label{RN-AdS}
RN-AdS_4: \qquad U^2(r)=1-\frac{2M}{r}+\frac{Q^2+P^2}{r^2}+g^2
r^2\ ,\qquad h(r)= r\ .
\end{equation}
Imposing BPS conditions leads to exactly two different possibilities in pure supergravity without vector
multiplets, as analyzed long ago in \cite{Romans:1991nq}. One solution is usually referred to as "extreme RN-AdS
electric solution", it is half-BPS and it requires $M=Q, P=0$, hence
\begin{equation}
extreme \quad electric \quad RN-AdS_4: \qquad
U^2(r)=(1-\frac{Q}{r})^2+g^2 r^2\, \qquad h(r) = r\ .
\end{equation}
The function $U(r)$ has no zeroes and therefore the spacetime has
no horizon. The point $r=0$ is then a naked singularity. The other
solution is referred to as an "exotic AdS solution" and is only
quarter-BPS, imposing $M = 0, P = 1/(2 g)$,
\begin{equation} exotic \quad AdS_4: \qquad
U^2(r)=(g r+\frac{1}{2 g r})^2+\frac{Q^2}{r^2}\ , \qquad h(r) = r\
.\end{equation} This case has no flat space limit for $g
\rightarrow 0$ and is therefore very different in behavior from
the first solution. Still, the solution has a naked singularity.

The aim of this paper is to find a generalization of the second solution within $N=2$ gauged supergravity with a
number of vector multiplets such that this naked singularity is resolved due to non-trivial scalar behavior. We
will focus on extending the exotic solution since the extension of the extreme RN-AdS solutions for many vector
multiplets and non-trivial scalars has been investigated in \cite{sabra-ads} with the outcome of nakedly
singular spacetimes once again. Some generalizations of the exotic solution also exist in the literature, e.g.
in \cite{sabra-cham}, but these set the scalars to constants and are thus not general enough to resolve the
naked singularity. Our strategy will be to replace the cosmological constant with a nontrivial potential for the
vector multiplet scalars that contains Fayet-Iliopoulos terms.

Anticipating our results, we now briefly explain how the exotic solution is modified to make proper black holes
in AdS$_4$. We set the electric charges to zero but allow for non-trivial scalars, which will in the end result
in changing the metric function $U$ to be\footnote{Here, the discussion is only schematic in order to underline
the main point, the actual solution is more involved as we explain in sections \ref{sect:4}-
\ref{sect:explcit_examples}. There we also comment further on the other function in the metric, $h(r)$.}:
\begin{equation}U^2(r)=(g r+\frac{c}{2 g r})^2\ ,\end{equation}
with a constant $c \neq 1$ that depends on the explicit running of
the scalars. The important outcome from this is that in certain
cases we will have $c < 0$, and then a horizon will appear at $r_h
= \sqrt{\frac{-c}{2 g^2}}$ to shield the singularity. In this way,
one can find a static quarter-BPS asymptotically AdS$_4$ black
hole with nontrivial scalar fields.

\section{Gauged supergravity with Fayet-Iliopoulos
parameters}\label{sect:2}

In this work we focus on abelian gauged $N=2$ supergravity in four dimensions in the absence of hypermultiplets.
We consider $n_V$ vector multiplets and keep the same conventions for metric signatures and field strenghts as
in \cite{Hristov:2009uj,our}. For some background material on gauged $N=2$ supergravity, see e.g.
\cite{deWit,DeWit:1984px,D'Auria:1990fj,Andrianopoli:1996cm}. As the gauge group is abelian, the vector
multiplet scalars are neutral, and the only charged fields in the theory are the two gravitinos. This is usually
referred to as Fayet-Iliopoulos (FI) gauging. The gauge fields that couple to the gravitinos appear in a linear
combination of the graviphoton and the $n_V$ vectors from the vectormultiplets, $\xi_{\Lambda}
A^{\Lambda}_{\mu}$, with $\Lambda = 0, 1,..., n_V$. The constants $\xi_{\Lambda}$ are called FI
parameters\footnote{\label{foot}The FI terms may also be understood from the triplet of quaternionic moment maps
$P^x_{\Lambda}$ in the absence of hypermultiplets. Using the local $SU(2)_R$ symmetry, we can rotate them such
that  $P^x_{\Lambda}= \delta^{x,3} \xi_{\Lambda}$, leaving a $U(1) \subset SU(2)_R$ as a residual symmetry. One
often uses the terminology that this part of the $R$-symmetry group is gauged.}. The bosonic part of the
Lagrangian for such a system is
\begin{align}\label{lagr}
\mathcal L=\frac{1}{2}R(g)+g_{i\bar \jmath}\partial^\mu z^i
\partial_\mu
{\bar z}^{\bar \jmath} +
I_{\Lambda\Sigma}F_{\mu\nu}^{\Lambda}F^{\Sigma\,\mu\nu}
+\frac{1}{2}R_{\Lambda\Sigma}\epsilon^{\mu\nu\rho\sigma}
F_{\mu\nu}^{\Lambda}F^{\Sigma}_{\rho\sigma} - g^2 V (z, \bar{z})\ ,
\end{align}
where
\begin{equation}\label{pot}
V= (g^{i\bar \jmath}f_i^\Lambda {\bar f}_{\bar \jmath}^\Sigma
-3{\bar L}^\Lambda L^\Sigma)\xi_\Lambda \xi_{\Sigma}
\end{equation}
is the scalar potential. Here, the complex scalar fields $z^i$ ($i
= 1,..., n_V$) are expressed in terms of holomorphic symplectic sections
$(X^{\Lambda} (z) ,F_{\Lambda}(z) )$ (see \cite{Andrianopoli:1996cm} for a review), and the
matrices $R_{\Lambda\Sigma}$ and $I_{\Lambda\Sigma}$ are the real and imaginary parts,
respectively, of the period matrix defined by
\begin{equation}\label{period-matrix}
{\overline {\cal N}}_{\Lambda \Sigma}\equiv \begin{pmatrix} D_iF_\Lambda \\
{\bar F}_\Lambda\end{pmatrix} \cdot {\begin{pmatrix} D_i X^\Sigma \\
{\bar X}^\Sigma\end{pmatrix}}^{-1} \ ,
\end{equation}
with $D_i \equiv (\partial_i+{\cal K}_i)$\footnote{More
explicitly, the period matrix can be computed by
$$
{\cal N}_{\Lambda \Sigma} = \overline{F}_{\Lambda \Sigma} + 2 i
\frac{{\rm Im}(F_{\Lambda \Gamma}) X^{\Gamma} {\rm Im}(F_{\Sigma
\Delta}) X^{\Delta} }{X^{\Gamma} {\rm Im}(F_{\Gamma \Delta})
X^{\Delta}}\ , \qquad F_{\Gamma \Delta} \equiv \frac{\partial
F_{\Gamma}}{\partial X^{\Delta}}\ .$$}. The K\"{a}hler potential
\begin{equation}\label{K-pot}
{\cal K}(z,\bar z)=-\ln\Big[i({\bar X}^\Lambda(\bar
z)F_\Lambda(z)-X^\Lambda(z) {\bar F}_\Lambda(\bar z))\Big]\
\end{equation}
determines the metric of the scalar field moduli space $g_{i\bar \jmath} = \partial_{z^i} \partial_{{\bar
z}^{\bar \jmath}} {\cal K} \equiv {\cal K }_{i\bar \jmath}$. In case a prepotential exists, it is given by
$F_\Lambda = \partial F/\partial X^\Lambda$, which we use in the examples discussed in section
\ref{sect:explcit_examples}. We will further make use of the quantities
\begin{equation}
(L^{\Lambda}, M_{\Lambda})
\equiv e^{{\cal K}/2} (X^{\Lambda}, F_{\Lambda})\ ,\qquad
(f^{\Lambda}_i, h_{\Lambda, i}) \equiv e^{{\cal K}/2} (D_i
X^{\Lambda}, D_i F_{\Lambda})\ .
\end{equation}
The supersymmetry variations for the gaugino and gravitino fields,
respectively, are:
\begin{align}
\delta_\varepsilon\lambda^{iA}&=i\partial_\mu z^i
\gamma^\mu\varepsilon^A - g^{i \bar{j}}
\bar{f}^{\Lambda}_{\bar{j}} I_{\Lambda \Sigma}
F^{\Sigma\,-}_{\mu\nu} \gamma^{\mu\nu}\epsilon^{AB}\varepsilon_B+i
gg^{i\bar \jmath}{\bar f}^\Lambda_{\bar \jmath}\xi_\Lambda
\sigma^{3, AB}\varepsilon_B\ ,\label{susygluino}
\\
\delta_\varepsilon \psi_{\mu A}&=\nabla_\mu\varepsilon_A +
2iF^{\Lambda\,-}_{\mu\nu}\,I_{\Lambda\Sigma} L^\Sigma \gamma^\nu
\epsilon_{AB}\varepsilon^B - \frac{g}{2} \sigma^3_{A B}
\xi_{\Lambda} L^{\Lambda} \gamma_\mu\varepsilon^B\
,\label{susy-gravi}
\end{align}
up to higher order terms in the fermions. This is sufficient for
solutions where all fermions are set to zero. The upper index "-"
on the fields strengths denotes their antiselfdual part. The
supercovariant derivative of the spinor reads:
\begin{equation}\label{supercovar_der}
    \nabla_\mu\varepsilon_A = (\partial_\mu - \frac{1}{4} \omega^{a b}_{\mu} \gamma_{a
    b}) \varepsilon_A + \frac{1}{4} \left({\cal K}_i\partial_\mu
z^i-{\cal K}_{\bar\iota}\partial_\mu{\bar z}^{\bar \iota} \right)
\varepsilon_A + \frac{i}{2} g \xi_{\Lambda} A^{\Lambda}_{\mu}
\sigma^3_A{}^B \varepsilon_B\ ,
\end{equation}
and similarly for the gravitino's
\begin{equation}\label{supercovar_der-gravitino}
    \nabla_\mu\psi_{\nu\,A} = \partial_\mu \psi_{\nu\,A}
   +...+ \frac{i}{2} g \xi_{\Lambda} A^{\Lambda}_{\mu}
\sigma^3_A{}^B \psi_{\mu\,B}\ .
\end{equation}
The fact that only $\sigma^3$ appears in the supersymmetry transformation rules and covariant derivatives
reflects the fact that the $SU(2)_R$ symmetry is broken to $U(1)$, as referred to in footnote \ref{foot}.

We have to stress that the above theory is gauged only electrically,
since we have used only electric fields $A_{\mu}^{\Lambda}$ for the gauging of the
gravitino. Thus the FI parameters can be thought of as the
electric charges $\pm e_\Lambda$ of the gravitino fields, with
\begin{equation}\label{gravitino-charges}
e_\Lambda =  g\xi_\Lambda\ ,
\end{equation}
The fact that the gravitinos have opposite electric charge finds its origin from the eigenvalues of $\sigma^3$.
Generically in such a theory one encounters a Dirac-like quantization condition in the presence of magnetic
charges $p^{\Lambda}$,
\begin{equation}\label{dir_quantization}
  2 e_{\Lambda} p^{\Lambda} = n\ , \qquad n \in \mathbb{Z}\ ,
\end{equation}
as explained in more detail in \cite{Romans:1991nq}. Clearly,
\eqref{dir_quantization} is not a symplectic invariant, due to the choice of the gauging.
Later, in section 6, we generalize this to include also magnetic gaugings.

\section{Black hole ansatz and Killing spinors}\label{sect:3}

As already stated in section 2, we look for a supersymmetric solution
similar to the "exotic AdS solution" of \cite{Romans:1991nq}, but
with nonconstant scalar fields. We start with the general static
metric ansatz
\begin{equation}\label{metric-ansatz} {\rm d} s^2 = U^2(r)\, {\rm
d}t^2 - U^{-2}(r)\, {\rm d}r^2 - h^2(r)\, ({\rm d} \theta^2 +
\sin^2 \theta {\rm d} \varphi^2)\ ,
\end{equation}
and corresponding vielbein
\begin{equation}
e_{\mu}^a = {\rm diag}\Big(U(r), U^{-1}(r), h(r), h(r) \sin
\theta\Big)\ .
\end{equation}
The non-vanishing components of the spin connection turn out to
be:
\begin{equation}
\omega_t^{0 1} = U \partial_r U, \quad
\omega_{\theta}^{1 2} = - U \partial_r h, \quad
\omega_{\varphi}^{13} = - U
\partial_r h \sin \theta, \quad \omega_{\varphi}^{23} = - \cos
\theta\ .
\end{equation}
We further assume that the gauge field
strengths are given by
\begin{equation}\label{el_field_strengths}
F^{\Lambda}_{t r} = 0, \qquad F^{\Lambda}_{\theta \varphi}
=\frac{p^{\Lambda}}{2} \sin \theta,
\end{equation}
or alternatively
\begin{equation}\label{gauge_fields}
A^{\Lambda}_t =  A^{\Lambda}_r =
A^{\Lambda}_{\theta} = 0, \qquad A^{\Lambda}_{\varphi} = -
p^{\Lambda} \cos \theta,
\end{equation}
which are needed in the BPS equations below. If we allow also electric charges, we then
should use an electromagnetic basis $F_{\mu \nu}^{\Lambda},
G_{\mu \nu, \Lambda}$\footnote{The magnetic field strengths can be
defined from the Lagrangian to be $$G_{\Lambda}{}_{\mu \nu} \equiv
R_{\Lambda
    \Sigma} F^{\Sigma}_{\mu \nu} - \frac 12 I_{\Lambda \Sigma}\,
  \epsilon_{\mu \nu \gamma \delta} F^{\Sigma \gamma \delta}\ .$$}, and require
\begin{equation}\label{solve_field_strengths}
G_{\Lambda, \theta \varphi} = \frac{q_{\Lambda}}{2} \sin \theta,
\qquad F^{\Lambda}_{\theta \varphi} = \frac{p^{\Lambda}}{2} \sin
\theta\ .
\end{equation}
These automatically solve the Maxwell equations and Bianchi
identities in full analogy to the case of ungauged
supergravity\footnote{Notice that the vector field part of the
Lagrangian \eqref{lagr} is the same as in the ungauged
theory, so they have the same equations of motion.The only difference appears in the coupling to the gravitinos.} \cite{Behrndt:1997ny}. However, we start with a purely electric
gauging \eqref{lagr} and we set the electric charges of the black hole to zero since
otherwise we cannot directly solve for the gauge fields
$A_t^{\Lambda}$ that are needed for the BPS equations. This is a
particular choice we make at this point in view of the BPS
conditions we derive below. In section \ref{electr-magn general
solution} we will explain how to explicitly find a solution also
with electric charges in a more general electromagnetic gauging
frame.

\subsection{Killing spinor ansatz}

With the gamma matrix conventions spelled out in Appendix \ref{app:A}
we make the following ansatz for the (chiral) Killing spinors:
\begin{equation}\label{KS-ansatz}
\varepsilon_A =  e^{i \alpha}\, \epsilon_{A B} \gamma^0
\varepsilon^B, \qquad \varepsilon_A =  \pm e^{i
\alpha}\,{\sigma}^3_{A B}\,\, {\gamma}^1\, \varepsilon^B\ ,
\end{equation}
where $\alpha$ is an arbitrary constant phase, and the choice of
sign in the second condition will lead to two distinguishable
Killing spinor solutions with corresponding BPS equations. This Killing spinor ansatz corresponds
(in our conventions for chiral spinors) to the Killing spinor
projections derived in \cite{Romans:1991nq} for the exotic
solutions. Note that the choice of phase $\alpha$ is irrelevant
due to $U(1)_R$ symmetry, i.e. any value of $\alpha$
leads to the exact same physical solution. It will nevertheless
amount to putting the symplectic sections of the vector multiplet
moduli space in a particular frame, as we explain in more
detail in the next subsection. Furthermore, from the above
equations one can deduce that the Killing spinor can be
parametrized as follows. Using our convention from App. \ref{app:A},
one finds that, $\forall a \in \mathbb{C}$, for the upper sign
(which we call type $I$) in \eqref{KS-ansatz}:
\begin{equation}
\varepsilon_1^I = a(x) \left(
\begin{array}{c}
1\\
i \\
-i \\
-1
 \end{array} \right), \qquad \varepsilon_2^I =
{\bar a(x)} e^{i \alpha}\,\left(
\begin{array}{c}
-i \\
1\\
1\\
-i
 \end{array} \right).
\end{equation}
For the negative sign (type II) one finds,
\begin{equation}
\varepsilon_1^{II} = a(x) \left(
\begin{array}{c}
1\\
i \\
i \\
1
 \end{array} \right), \qquad \varepsilon_2^{II} =
{\bar a(x)} e^{i \alpha}\,\left(
\begin{array}{c}
i \\
-1\\
1\\
-i
 \end{array} \right).
\end{equation}
This type of Killing spinors explicitly break $3/4$ of the
supersymmetry. The two degrees of freedom of the complex function
$a$ give the remaining two supercharges.

We look for spacetimes that are static and spherically symmetric,
so in particular invariant under the rotation group. This rotation
group acts on spinors, and can in general leave or not leave our
Killing spinor ansatz invariant. It will be a check on our
explicit solution for the Killing spinors that they should be also
rotationally invariant, just as in the original case for exotic
solutions \cite{Romans:1991nq}.

Note that our choice of Killing spinors makes them timelike, i.e. they give rise to a timelike Killing vector
(see \cite{Kallosh:1993wx,ortinetal} for more details about Killing spinor identities). One can then show
\cite{our} that,  to obtain a supersymmetric solution, one needs to check only the Maxwell equations and Bianchi
identities in addition to the BPS conditions. The equations of motion for the other fields then follow, due to
the timelike Killing spinor.

\subsection{BPS conditions and attractor flow}

With the above ans\"{a}tze for the spacetime and the Killing spinors one can show that  the gaugino and
gravitino variations \eqref{susygluino}, \eqref{susy-gravi} simplify substantially but do not yet vanish
identically.

From the gaugino variation we obtain the following radial flow equations for the scalar fields:
\begin{equation}\label{gaugino-var}
e^{-i \alpha} U \partial_r z^i = g^{i \bar{j}}
\bar{f}^{\Lambda}_{\bar{j}} \left( \frac{2 I_{\Lambda \Sigma}
p^{\Sigma}}{h^2} \mp g \xi_{\Lambda}
 \right),
\end{equation}
where the two different signs correspond to the two types of
Killing spinors in the given order.

If we require the gravitino variation \eqref{susy-gravi} to vanish, we derive four extra equations that need to
be satisfied (one for each spacetime index). The equations for $t$ and $\theta$ determine the radial dependence
of the metric components,
\begin{equation}\label{gravitino-var1}
e^{i \alpha} \partial_r U = - \frac{2 L^{\Lambda} I_{\Lambda
\Sigma} p^{\Sigma}}{h^2} \pm g \xi_{\Lambda} L^{\Lambda},
\end{equation}
\begin{equation}\label{gravitino-var2}
e^{i \alpha} \frac{U}{h} \partial_r h  = \frac{2 L^{\Lambda}
I_{\Lambda \Sigma} p^{\Sigma}}{h^2} \pm g \xi_{\Lambda}
L^{\Lambda}\ .
\end{equation}
The $\varphi$ component of the gravitino variation further constrains
\begin{equation}\label{gravitino-var3}
2 g \xi_{\Lambda} p^{\Lambda} = \mp 1\ ,
\end{equation}
and the radial part gives a differential equation for the Killing spinor, solved by
\begin{equation}\label{gravitino-var4}
a(r) = a_0\ \sqrt{U(r)}\ e^{-\frac{i}{2}\int A_r(r)\, {\rm d} r}\
,
\end{equation}
with
\begin{equation}\label{K-connect}
A_r(r) =-\frac{i}{2} \left(\mathcal{K}_i \partial_r z^i -
\mathcal{K}_{\bar{j}} \partial_r z^{\bar{j}} \right)
\end{equation}
the $U(1)$ K\"{a}hler connection. These results are in agreement
with rotational symmetry since the Killing spinor is only a
function of $r$. The solution is 1/4 BPS and has two conserved
supercharges, corresponding to the two free numbers of the complex
constant $a_0$. We further see that \eqref{gravitino-var4} does
not give an extra constraint on the fields, but can be used to
determine the explicit radial dependence of the Killing spinor
parameter $a(r)$. One can always evaluate the integral of $A(r)$
for a given solution and thus the Killing spinor can be explicitly
found once the BPS equations
\eqref{gaugino-var}-\eqref{gravitino-var3} are satisfied.

Notice also that \eqref{gravitino-var3} is in accordance with the generalized Dirac quantization condition
\eqref{dir_quantization} with the smallest non-zero integer $n=\pm 1$. It will be interesting to understand how
one can generate other solutions with higher values of $n$ or whether supersymmetry always strictly constrains
$n$ as in the present case. Furthermore, it is easy to see that in the limit $g \rightarrow 0$ where the gauging
vanishes one recovers the well-known first order attractor flow equations of black holes in ungauged $N=2$
supergravity \cite{Ferrara:1995ih,Strominger:1996kf,Ferrara:1996dd}. The presence of the extra terms due to the
gauging is precisely where the difference between ungauged and gauged black holes lies. Thus we believe the BPS
equations are now written in a simpler and more suggestive form compared to \cite{klemm-adsBH}.

A short comment on the phase $\alpha$ is in order. One can see in
eqs. \eqref{gravitino-var1}, \eqref{gravitino-var2} that the
quantities $e^{-i \alpha} L^{\Lambda}$ must always be real. Thus,
if e.g. $\alpha = 0$ then $L^{\Lambda}$ will need to be real,
while if $\alpha = \frac{\pi}{2}$, $L^{\Lambda}$ have to be
imaginary. This $U(1)_R$ symmetry of the BPS conditions is of
course well understood in the ungauged case and there are
generally two ways of proceeding. One can just fix the phase to a
particular value and go on to write down the solutions, as
originally done in \cite{Behrndt:1997ny}, or one can also put
explicitly the phase factor in the definition of the sections as
done in \cite{Denef:2000nb}. Here we choose to fix $\alpha = 0$
for the rest of the paper as it will minimize the factors of $i$
in what follows (note that \cite{Behrndt:1997ny} makes the
opposite choice and thus the solutions are given for the imaginary
instead of the real parts of the sections). It should be clear how
one can always plug back the factor of $e^{-i \alpha}$ and choose
a different phase if needed in different conventions. In
particular this choice implies that (after adding
\eqref{gravitino-var1} and \eqref{gravitino-var2})
\begin{equation}\label{imag_sections}
\xi_{\Lambda} {\rm Im}(X^{\Lambda}) = 0\ .
\end{equation}

\section{Black hole solutions}\label{general_solution}\label{sect:4}

Now we would like to find explicit solutions to eqs. \eqref{gaugino-var}-\eqref{gravitino-var2}. We already know
(by assumption) the solution for the vector field strengths \eqref{el_field_strengths}, so we search for
solutions of the metric functions $U(r), h(r)$ and the symplectic sections $X^{\Lambda} (r), F_{\Lambda} (r)$
that determine the scalars. We propose the following form for the solution of the BPS equations in the electric
frame (for the choice of phase $\alpha = 0$):
\begin{equation}\label{scalar_sol}
 \frac{1}{2} \left(X^{\Lambda} + \bar{X}^{\Lambda} \right) = H^{\Lambda}\ , \qquad \frac{1}{2} \left(F_{\Lambda} + \bar{F}_{\Lambda} \right) = 0\ ,
\end{equation}
$$H^{\Lambda} = \alpha^{\Lambda} + \frac{\beta^{\Lambda}}{r},$$
and
\begin{equation}\label{metric_sol}
  U(r) = e^{\mathcal{K}/2} \left(g r + \frac{c}{2 g r} \right)\ , \qquad h(r) = r e^{-
  \mathcal{K}/2}\ ,
\end{equation}
where $\mathcal{K}$ is the K\"{a}hler potential
\begin{equation}\label{Kaehler}
    e^{- \mathcal{K}} = i \left(
  \bar{X}^{\Lambda} F_{\Lambda} - X^{\Lambda} \bar{F}_{\Lambda}
  \right)\ ,
\end{equation}
and $c$ some constant. The line element of the spacetime is then
\begin{equation}\label{line-element_sol}
  {\rm d} s^2 = e^{\mathcal{K}} \left(g r+\frac{c}{2 g r} \right)^2 {\rm d}t^2 - \frac{e^{-\mathcal{K}} {\rm d}r^2}{\left(g r+\frac{c}{2 g r} \right)^2} - e^{-\mathcal{K}} r^2 {\rm d} \Omega_2^2\ .
\end{equation}
The constant $c$ above is not specified yet and depends explicitly
on the chosen model. This is also the case for the constants
$\alpha^{\Lambda}, \beta^{\Lambda}$ that may eventually be
expressed in terms of the FI parameters $\xi_{\Lambda}$ and the
magnetic charges $p^{\Lambda}$. We give some explicit examples in
section \ref{sect:explcit_examples}. Here we just use the above
results to show how the BPS equations simplify to a form where
they can be explicitly solved given a particular model with a
prepotential (we further assume that \eqref{scalar_sol} implies
${\rm Im}(X^{\Lambda}) = 0$ in accordance with
\eqref{imag_sections}). Eqs.
\eqref{gravitino-var1}-\eqref{gravitino-var2}, together with
\eqref{scalar_sol}-\eqref{metric_sol}, lead to:
\begin{equation}\label{final_eq1}
  \xi_{\Lambda} \alpha^{\Lambda} = \pm 1, \qquad \xi_{\Lambda}
  \beta^{\Lambda} = 0\ ,
\end{equation}
\begin{equation}\label{final_eq2}
  F_{\Lambda} \left( -2 g^2 r \beta^{\Lambda} + c
  \alpha^{\Lambda}+2 g p^{\Lambda}\right)= 0\ .
\end{equation}

Multiplying \eqref{gaugino-var} with $f_i^{\Lambda}$ we eventually
obtain
\begin{align}\label{final_eq3}
\begin{split}
\left( g r + \frac{c}{2 g r} \right) \left( F_{\Sigma} X^{\Sigma} \partial_r X^{\Lambda} - X^{\Lambda} F_{\Sigma} \partial_r  X^{\Sigma} \right) &= -\frac{1}{r^2} F_{\Sigma} \left(X^{\Sigma} p^{\Lambda} - X^{\Lambda}  p^{\Sigma} \right) \\
& + g F_{\Sigma} X^{\Sigma} \left( X^{\Lambda} \pm i F_{\Pi}
X^{\Pi} (I^{-1})^{\Lambda \Gamma} \xi_{\Gamma} \right)\ .
\end{split}
\end{align}
We chose to rewrite it in this form in order to have equations
only for the symplectic sections, as standardly done also in
ungauged black holes literature \cite{Behrndt:1997ny}. In
principle however $f_i^{\Lambda}$ is non-invertible and thus
\eqref{final_eq3} does not strictly speaking imply
\eqref{gaugino-var}. Practically this never seems to be an issue since in
fact \eqref{final_eq3}  gives one extra equation. In all cases we solved explicitly the
equations, we found that the condition coming from the gaugino
variation is already automatically satisfied after solving
\eqref{final_eq1} and \eqref{final_eq2}. Unfortunately, we were
not able to prove that it must vanish identically with the above
ansatz.

Using \eqref{scalar_sol} it is straightforward to prove that the
K\"{a}hler connection \eqref{K-connect} vanishes identically (c.f.
Eq.(29) of \cite{Behrndt:1997ny}). Thus the functional dependence
of the Killing spinors becomes
\begin{equation}\label{gravitino-var4}
a(r) = \sqrt{U(r)}\ a_0\ ,
\end{equation}
just as in the original solution without scalars
\cite{Romans:1991nq}.

Note that with \eqref{scalar_sol} one can now also show that the
field strengths \eqref{el_field_strengths} identically solve the
Bianchi identities and the Maxwell equation as they fall in the
form \eqref{solve_field_strengths} with $q_{\Lambda} = 0$. Thus
any solution of \eqref{final_eq1}-\eqref{final_eq3} will be a
supersymmetric solution of the theory with no further constraints.

One particular solution (the only one in absence of vector multiplets) of the above equations that is always
present, is when $\alpha^{\Lambda} = -2 g p^{\Lambda}$, $\beta^{\Lambda} = 0$, for all $\Lambda$, and $c = 1$.
This solution is in fact the one discovered in \cite{sabra-cham} with constant scalars ($X^{\Lambda}$ is
constant when $\beta^{\Lambda} = 0$). However, this solution has a naked singularity, since $c>0$. A horizon is
not present in this case,  since generally it will appear at $r_h^2 = -\frac{c}{2 g^2}$ and thus only for $c <
0$. We will see in section \ref{sect:explcit_examples} that indeed there exist solutions of the above equations
in which $c<0$, such that a proper horizon shields the singularity. These solutions however necessarily have
nonzero $\beta^{\Lambda}$'s. Thus a proper black hole can only form in the presence of some sort of attractor
mechanism for the scalar fields.

\section{Black holes with electric and magnetic
charges}\label{electr-magn general solution}

We now explain how one can restore the broken electromagnetic duality invariance of the theory \eqref{lagr}. As
discussed in section \ref{sect:2}, the electric gaugings break electromagnetic invariance, i.e. performing
symplectic rotations leads us to a new Lagrangian that will be of different form from \eqref{lagr}. One then
needs to allow for both electric and magnetic gaugings and change the form of the scalar potential in order to
recover the electromagnetic invariance of the ungauged theory. There have been various proposals in literature
for extending it to gauged supergravity \cite{Michelson,tens1}. It turns out that the correct approach to
introducing real magnetic gaugings is the embedding tensor formalism, and we closely follow the analysis of
\cite{embedding}. It restores full electromagnetic duality invariance of the gauge theory (when the electric and
magnetic charges are mutually local) by introducing additional tensor fields in the Lagrangian. Unfortunately
the theory is not yet fully developed in general for supergravity (for rigid $N=2$ supersymmetry, see \cite{de
Vroome:2007zd}), but we will nevertheless be able to write down particular solutions due to the fact that we can
do duality transformations on the solutions of the electrically gauged theory.

Even though we cannot give the most general Lagrangian and susy variations for the theory with electric and
magnetic gaugings, we know how the bosonic part of the Lagrangian should look like in this very special case of
FI gaugings. It is most instructive to integrate out the additional tensor field that has to be introduced,
following the procedure of section 5.1 of \cite{embedding}. Exactly half of the gauge fields (we will originally
have both electric and magnetic gauge fields, $(A_{\mu}^{\Lambda}, A_{\mu, \Lambda})$) will also be integrated
out in this process. One first splits the index $\Lambda$ in two parts, $\{ \Lambda \} = \{ \Lambda',\Lambda''
\}$, for the non-vanishing electric and magnetic gauge fields respectively. The Lagrangian will then consist
only of $A_{\mu}^{\Lambda'}, A_{\Lambda'', \mu}$, while $A_{\mu}^{\Lambda''}, A_{\Lambda', \mu}$ are integrated
out together with the additional tensor field. Thus the linear combination of fields used for the $U(1)$ FI
gauging is $\xi_{\Lambda'} A_{\mu}^{\Lambda'} - \xi^{\Lambda''} A_{\Lambda'', \mu}$. The $\xi^{\Lambda''}$'s are
the magnetic charges of the gravitinos, and the new generalized Dirac quantization condition for electric and
magnetic charges $(q_{\Lambda},p^{\Lambda})$ of any solution is
\begin{equation}\label{dir_quantization_em}
  2 ( e_{\Lambda'} p^{\Lambda'} - m^{\Lambda''} q_{\Lambda''} ) = n, \qquad n \in \mathbb{Z}\ ,
\end{equation}
with electric and magnetic gravitino charges $e_{\Lambda'} \equiv g \xi_{\Lambda'}$ and $m^{\Lambda''} \equiv g
\xi^{\Lambda''}$. The scalar potential is then of the form
\begin{equation}\label{pot_em}
V= (g^{i\bar \jmath}f_i^{\Lambda'} {\bar f}_{\bar \jmath}^{\Sigma'} -3{\bar L}^{\Lambda'}
L^{\Sigma'})\xi_{\Lambda'} \xi_{\Sigma'} - (g^{i\bar \jmath} h_{i, \Lambda''} {\bar h}_{\bar{\jmath}, \Sigma''}
-3{\bar M}_{\Lambda''} M_{\Sigma''})\xi^{\Lambda''} \xi^{\Sigma''}\ .
\end{equation}
The main point about electromagnetic invariance is that the equations of motion are now invariant under the
group $Sp(2 (n_V+1), \mathbb{R})$, which at the same time rotates the Lagrangian from a purely electric gauging
frame to a more general electromagnetic gauging. The symplectic vectors transforming under the symmetry group
are the sections $(F_{\Lambda}, X^{\Lambda})$ and the FI parameters $(\xi_{\Lambda}, \xi^{\Lambda})$, as well as
the vector field strengths $F_{\mu \nu}^{\Lambda}, G_{\mu \nu, \Lambda}$ (which come from the respective
electric and magnetic gauge potentials $(A_{\mu}^{\Lambda}, A_{\mu, \Lambda})$). One can then see how natural
equations \eqref{dir_quantization_em},\eqref{pot_em} are if we start from a purely electric frame with only
$\xi_{\Lambda}, F_{\mu \nu}^{\Lambda}$ nonzero and then perform an arbitrary symplectic transformation. The
important message is that once we have found a solution to the purely electric theory we can always perform any
symplectic transformation of the theory to see how the solution looks like in a more general electromagnetic
setting.

It is in fact easy to guess how the solution looks like in a more general theory with electric and magnetic
gaugings. We have not proven the existence of such a BPS solution due to the lack of a properly defined
Lagrangian and supersymmetry variations, but we can nevertheless indirectly find it by symplectic rotations.
This procedure leads to  a solution, where the metric is again given by  \eqref{line-element_sol}, together with
\begin{align}
\begin{split}
F^{\Lambda'}_{t r} = 0\ , \qquad F^{\Lambda'}_{\theta \varphi} =
\frac{p^{\Lambda'}}{2} \sin \theta\ ,\\
G_{\Lambda'', t r} = 0\ , \qquad G_{\Lambda'',\theta \varphi} =
\frac{q_{\Lambda''}}{2} \sin \theta\ ,
\end{split}
\end{align}
and harmonic functions that determine the sections
\begin{align}
\begin{split}
\frac{1}{2} \left(X^{\Lambda'} + \bar{X}^{\Lambda'} \right) =
H^{\Lambda'}\ , \qquad \frac{1}{2} \left(F_{\Lambda'} +
\bar{F}_{\Lambda'} \right) = 0\ ,
\\
\frac{1}{2} \left(X^{\Lambda''} + \bar{X}^{\Lambda''} \right) = 0\
, \qquad \frac{1}{2} \left(F_{\Lambda''} + \bar{F}_{\Lambda''}
\right) = H_{\Lambda''}\ ,
\end{split}
\end{align}
$$H^{\Lambda'} = \alpha^{\Lambda'} + \frac{\beta^{\Lambda'}}{r}\ , \qquad H_{\Lambda''} = \alpha_{\Lambda''} + \frac{\beta_{\Lambda''}}{r}\ .$$
The above should give solutions provided that the following
identities (coming from the BPS conditions) are satisfied,
\begin{equation}\label{eq0_em}
  2 g ( \xi_{\Lambda'} p^{\Lambda'} - \xi^{\Lambda''} q_{\Lambda''}) = \mp
  1\ ,
\end{equation}
\begin{equation}\label{eq1_em}
  \xi_{\Lambda'} \alpha^{\Lambda'} - \xi^{\Lambda''} \alpha_{\Lambda''} = \pm 1, \qquad \xi_{\Lambda'}
  \beta^{\Lambda'} - \xi^{\Lambda''} \beta_{\Lambda''} = 0\ ,
\end{equation}
\begin{equation}\label{eq2_em}
  F_{\Lambda'} \left( -2 g^2 r \beta^{\Lambda'} + c
  \alpha^{\Lambda'}+2 g p^{\Lambda'}\right) - X^{\Lambda''} \left( -2
  g^2 r \beta_{\Lambda''} + c \alpha_{\Lambda''}+2 g q_{\Lambda''}
\right)= 0\ ,
\end{equation}
together with the symplectic invariant version of
\eqref{final_eq3} coming from contraction with $f_i^{\Lambda}$.
This expression becomes lengthy and cumbersome to check
and we will not write it down explicitly. In this case it will be
easier to explicitly check the symplectic invariant version of
\eqref{gaugino-var} by first defining the complex vector multiplet
scalars from the sections. Of course in case of confusion one can
always take a model and rotate it to the electric frame where the
susy variations are clearly spelled out
(\eqref{susygluino}-\eqref{susy-gravi}).

\section{Explicit black hole solutions}\label{sect:explcit_examples}

\subsection{$n_V = 1$ with $F =-2 i \sqrt{X^0 (X^1)^3}$}\label{sect:5.1}
This is the simplest prepotential in the ordinary electrically
gauged theory that leads to a black hole solution. We have one
vector multiplet with the prepotential
\begin{equation}F =-2 i \sqrt{X^0 (X^1)^3}\ ,\end{equation}
thus one finds $X^0 = \alpha^0+\frac{\beta^0}{r}, X^1 =
\alpha^1+\frac{\beta^1}{r}$ from \eqref{scalar_sol}. This theory
exhibits an AdS$_4$ vacuum at the minimum of the scalar potential
(corresponding to the cosmological constant)
\begin{equation}
V^* = \Lambda = - \frac{2 g^2}{\sqrt{3}} \sqrt{\xi_0 \xi_1^3}
\end{equation}
at $z^* = \sqrt{\frac{3 \xi_0}{\xi_1}}$ (defining $z \equiv
\frac{X^1}{X^0}$). This can be easily deduced using the results of
\cite{Hristov:2009uj}. Going through the BPS equations
\eqref{final_eq1}-\eqref{final_eq2}, we can fix all the constants
of the solution in terms of the FI parameters $\xi_0, \xi_1$ apart
from one free parameter (here we leave $\beta^1$ to be free for
convenience, but it can be traded for one of the magnetic charges
or for $\beta^0$). We obtain that the magnetic charges are given
by:
\begin{equation}\label{magn-charges}
p^0 = \mp \frac{1}{g \xi_0} \left(\frac{1}{8}+\frac{8 (g \xi_1 \beta^1)^2}{3} \right), \quad p^1 = \mp \frac{1}{g \xi_1} \left(\frac{3}{8}-\frac{8 (g \xi_1 \beta^1)^2}{3} \right)\ ,
\end{equation}
for spinor I and II respectively. The other constants in the
solution are
\begin{equation}\label{constants}
\beta^0 = -\frac{\xi_1 \beta^1}{\xi_0}, \qquad \alpha^0 = \frac{\pm 1}{4 \xi_0}, \qquad \alpha^1 = \frac{\pm 3}{4 \xi_1}, \qquad c = 1 - \frac{32}{3} (g \xi_1 \beta^1)^2\ .
\end{equation}
Using the definition of the gravitino charges \eqref{gravitino-charges}, $e_\Lambda = g \xi_\Lambda$, these relations imply
\begin{equation}
e_\Lambda \alpha^\Lambda = \pm g\ , \qquad e_\Lambda \beta^\Lambda = 0\ , \qquad 2 e_\Lambda p^\Lambda = \mp 1\ ,
\end{equation}
and one can check that the complete solution is a function of the
variables $e_\Lambda, p^\Lambda$ and $g$. Note that in fact the
dependence on $g$ is artificial since it can always be absorbed in
the definition of the coordinates. In particular, the rescaling $g
r \rightarrow r, t \rightarrow g t$ makes the metric and the
scalar flow dependent only on $e_\Lambda, p^\Lambda$ as is also
the form of the solution presented in \cite{klemm-adsBH}.

Interestingly, one can verify that the condition coming from the
gaugino variation, \eqref{final_eq3}, is automatically satisfied
with no further constraints. One can see that the two spinor types
in the end amount to having opposite magnetic charges and to
flipping some signs for the solution of the sections.

We now
analyze the physical properties of the solution. In this case it
is important to give explicitly the metric function in front of
the ${\rm d} t^2$ term. Using the form of the line element in \eqref{line-element_sol}, the specific form of the sections with constants given in \eqref{constants}, one can explicitly compute:
\begin{equation}
g_{tt} = \frac{2 \sqrt{\xi_0 \xi_1^3} r^2 \left(g r+\frac{1}{2 g
r}-\frac{16 g}{3 r} (\xi_1 \beta^1)^2 \right)^2}{\sqrt{(r \mp 4
\xi_1 \beta^1)(3 r \pm 4 \xi_1 \beta^1)^3}}\ .
\end{equation}
The leading terms of the (infinite) asymptotic expansion of the
metric for $r\rightarrow\infty$ are then
\begin{equation}
g_{tt} (r \rightarrow \infty) = -\frac{\Lambda r^2}{3} \left( 1 +
\frac{1}{2 g^2} (1+c) \frac{1}{r^2} - \frac{256 (\xi_1
\beta^1)^3}{27} \frac{1}{r^3} + \mathcal{O} \left(\frac{1}{r^4}
\right) \right)\ .
\end{equation}
Clearly, the metric has the correct AdS$_4$ asymptotics. Although
the constant term of the asymptotic expansion is not exactly $1$
when we compare to the RN-AdS metric of section \ref{overview}, we
are still tempted to think that the coefficient in front of the
$1/r$ term determines the physical mass of the black hole,
\begin{equation}\label{proposedmass}
M = - \frac{128}{81} \Lambda (\xi_1 \beta^1)^3\ .
\end{equation}
The issue of defining the mass is a bit more subtle in
asymptotically AdS spacetimes and we address it more carefully in
section \ref{sect:Thermo}, where we verify our expectation.

One can also notice that there are some subtleties for the radial
coordinate that usually do not appear for black hole spacetimes.
In particular, $r = 0$ is neither a horizon (where $g_{tt} = 0$),
nor a singularity (where $g_{tt} \rightarrow \infty$). In fact the
point $r = 0$ is never part of the spacetime, since the
singularity is always at a positive $r$, where the space should be
cut off. Thus the $r$ coordinate does not directly correspond to
the radial coordinate from the singularity. The horizon for both
signs is at
\begin{equation}
r_h = \sqrt{\frac{16}{3} (\xi_1
\beta^1)^2 - \frac{1}{2 g^2}}\ ,
\end{equation}
while genuine singularities will appear at $r_s = \pm 4 \xi_1 \beta^1, \mp \frac{4}{3} \xi_1 \beta^1$. The
spacetime will then continue only until the first singularity is encountered. If we want to have an actual black
hole spacetime we must insist that the horizon shields the singularity, i.e. $r_h
> r_s$, otherwise we again have a naked singularity and the sphere at $r_h$ will not be part of the spacetime.
This requirement further sets the constraints $|g \xi_1 \beta^1| >
\frac{3}{8}$, with $\xi_1 \beta^1 < 0$ for solution I (upper sign) and $\xi_1
\beta^1 > 0$ for solution II (lower sign). Since the parameter $\beta^1$ is at
our disposal, it can always be chosen to be within the required
range, thus the singularity can be shielded by a horizon in a
particular parameter range for $\beta^1$. So, putting together
both solutions, we know that a proper black hole with a horizon
will form in case $g \xi_1 \beta^1 \in (-\infty,
-\frac{3}{8})\bigcup (\frac{3}{8}, \infty)$, with the
corresponding relations given above between the magnetic charges
and $\xi_1 \beta^1$ for the two intervals. In between, we are dealing with naked singularities, which are of no
interest for us at present. The constant $c$ is always negative, and satisfies
\begin{equation}
c < -\frac{1}{2}\ ,
\end{equation}
which reflects again the existence of a horizon, as announced in section 2.

Let us now investigate further the properties of
these new black holes. Their entropy is proportional to the area
of the black hole at the horizon,
\begin{equation}
    S = \frac{A}{4} = \frac{3}{4 \Lambda} \sqrt{(r_h-r_{s,1}) (r_h-r_{s,2})^3} = \frac{\sqrt{(r_h \mp 4 \xi_1 \beta^1)(3 r_h \pm 4 \xi_1 \beta^1)^3}}{8 \sqrt{\xi_0 \xi_1^3}}\ ,
\end{equation}
so the entropy is effectively a function of $\xi_0, \xi_1,
\beta^1$, which can be rewritten in terms of the FI-terms and
magnetic charges. Thus the entropy is a function of the black hole
charges $p^{\Lambda}$ and the gravitino charges $e_{\Lambda}$. One
can further observe that in case of fixed gravitino charges
$e_{\Lambda}$, the entropy scales quadratically with the parameter
$\beta^1$ and thus linearly with the charges $p^0$ or $p^1$ in the
limit of large charges. The opposite limit of fixed magnetic
charges shows that the entropy remains constant for large
gravitino charge.

It is interesting to note that the fact that the scalars at the
horizon are fixed in terms of the gravitino and black hole charges
is not directly obvious from the general form of the solution. The
scalars depend on the constants $\alpha^{\Lambda}, \beta^{\Lambda}
$ that might not always be fully determined by $\xi_{\Lambda},
p^{\Lambda}$. One example of this is for the prepotential $F = -i
X^0 X^1$ where the magnetic black hole charges are fully fixed in
terms of FI parameters and either $\beta^0$ or $\beta^1$ can be
freely chosen. However, one can show that in this case there is no
parameter range for the $\beta^{\Lambda}$'s where the singularity
is shielded by the horizon, thus black holes do not exist. In all
the cases for which we checked that a black hole is possible we
could verify that indeed the scalar values at the horizon can be
expressed in terms of the charges and FI parameters, but we have
no general proof of this\footnote{The BPS equations
\eqref{final_eq1}-\eqref{final_eq3} can be relatively easily
solved in full generality for a prepotential of the form $F =
(X^0)^n (X^1)^{2-n}$. The outcome is that black holes exist for $n
\in (0,1)$. The solution for general $n$ is in full analogy to the
one presented here. There is only certain $n$ dependence in the
way the various constants depend on each other, which does not
lead to any qualitative differences. Here we chose to explicitly
describe the case with $n=1/2$ since it is the most relevant case
from a string theory point of view as we will see in the next
section.}.

Another interesting question is what the near-horizon geometry of
this black hole is. It is natural to expect that a static four
dimensional BPS black hole has a near-horizon geometry of $AdS_2
\times S^2$ and this is indeed the case. The radii of the two
spaces are
\begin{equation}
R_{S^2} = r_h
e^{-\mathcal{K}/2}|_{r=r_h}, \qquad R_{AdS_2} =
\frac{e^{-\mathcal{K}/2}|_{r=r_h}}{2 \sqrt{2} g}\ ,
\end{equation}
and it can be shown that $R_{S^2} > \sqrt{2} R_{AdS_2}$ from the
constraints on having a horizon. As the radii are inversely
proportional to the scalar curvature of these spaces, it follows
that the overall $AdS_2 \times S^2$ space has a negative
curvature, as expected for asymptotically AdS$_4$ black holes.
Thus it is clear that near the horizon we do not observe a
supersymmetry enhancement to a fully BPS vacuum as is the case for
the asymptotically flat static BPS black holes\footnote{$AdS_2
\times S^2$ is maximally supersymmetric only for $R_{S^2} =
R_{AdS_2}$ as shown in \cite{Hristov:2009uj}}. Nevertheless, there
could be a supersymmetry enhancement from a 1/4 BPS overall
solution to a 1/2 BPS vacuum near the horizon.

\subsection{$F =\frac{(X^1)^3}{X^0}$ in a mixed electromagnetic frame}
In order to give an example of black hole solutions in a more
general electromagnetic frame, one can rotate the sections and FI
parameters of the previous example by the symplectic matrix
\begin{equation}
\mathcal{S}=\begin{pmatrix} 1 & 0 & 0 & 0 \\ 0 & 0 & 0 & 1/3 \\
0 & 0 & 1 & 0\\ 0 & -3 & 0 & 0 \end{pmatrix}\ ,
\end{equation} such
that the prepotential after rotation corresponds to the
well-studied in ungauged supergravity $T^3$ model with
prepotential
\begin{equation}
F =\frac{(X^1)^3}{X^0}\ ,
\end{equation}
and the non-vanishing FI
parameters are $\xi_0, \xi^1$. The theory will then be
electrically gauged with $A_{\mu}^0$ and magneticaly gauged with
$A_{1, \mu}$. This prepotential cannot lead to an AdS BPS black
hole in the purely electric gauging, because it does not exhibit a
supersymmetric AdS$_4$ vacuum, as one can find using the methods
of \cite{Hristov:2009uj}. However, in this mixed electromagnetic
gauging, the $T^3$ model does have a proper fully supersymmetric
AdS vacuum.

Now we can follow the more general procedure outlined in section
\ref{electr-magn general solution}. In this case it turns out that
$X^0 = \alpha^0+\frac{\beta^0}{r}, F_1 =
\alpha_1+\frac{\beta_1}{r}$. The black hole solution will then
have one magnetic charge $p^0$ and one electric charge $q_1$.
Going through the BPS equations \eqref{eq0_em}-\eqref{eq2_em}, we
can fix all the constants of the solution in terms of the FI
parameters $\xi_0, \xi^1$, apart from one free parameter which we
choose to be $\beta_1$. The charges are given by:
\begin{equation}
p^0 = \mp \frac{1}{g \xi_0} \left(\frac{1}{8}+\frac{8 (g \xi^1
\beta_1)^2}{3} \right), \quad q_1 = \pm \frac{1}{g \xi^1}
\left(\frac{3}{8}-\frac{8 (g \xi^1 \beta_1)^2}{3} \right)\ ,
\end{equation}
for spinor I and II respectively. The other constants in the
solution are \begin{equation}\beta^0 = \frac{\xi^1
\beta_1}{\xi_0}, \qquad \alpha^0 =\pm  \frac{1}{4 \xi_0}, \qquad
\alpha_1 = \mp \frac{ 3}{4 \xi^1}, \qquad c = 1 - \frac{32}{3} (g
\xi^1 \beta_1)^2\ .\end{equation} and one can see that the metric
and scalar profile in this case are analogous to the example in
the previous subsection, as expected. This confirms the
consistency of the results in section \ref{electr-magn general
solution}. The entropy of the black hole is now a function of the
electric and magnetic gravitino charges, $e_0 = g \xi_0$ and $m^1
= g \xi^1$, and the black hole charges $p^0$ and $q_1$.

Note that we could have for instance rotated the frame from a
fully electric to a fully magnetic frame, by the symplectic matrix
\begin{equation}
\mathcal{S}=\begin{pmatrix} 0 & 0 & -1 & 0 \\ 0 & 0 & 0 & -1/3 \\
1 & 0 & 0 & 0\\ 0 & 3 & 0 & 0 \end{pmatrix}\ ,
\end{equation} and
it turns out that the prepotential $F =-2 i \sqrt{X^0 (X^1)^3}$ is
in fact invariant under this transformation. The resulting
solution will be the same, but there will be two electric instead
of two magnetic charges.

\section{M-theory lift}\label{sect:M-theory reduction}

An explicit string theory example of abelian gauged $N=2$, $D=4$
supergravity with FI terms was found by a consistent truncation of
M-theory on $S^7$ in \cite{duff}. A standard Kaluza-Klein
compactification on $S^7$ leads initially to an $SO(8)$ gauged
$N=8$ supergravity in four dimensions. To avoid some of the
complications of non-abelian gauge fields, the authors of
\cite{duff} further defined a consistent truncation of this theory
to an $U(1)^4$ gauged $N=2$ supergravity. The 11-dimensional
metric ansatz is given by:
\begin{equation}\label{11dim metric}
    ds^2_{11} = \Delta^{2/3} ds_4^2 + 2 g^{-2} \Delta^{-1/3}
    \sum_{\Lambda = 0}^3 a_{\Lambda}^{-1} \left(d\mu_{\Lambda}^2 + \mu_{\Lambda}^2 (d\phi_{\Lambda}+\frac{g}{\sqrt{2}} A^{\Lambda})^2
    \right),
\end{equation}
where $\Delta = \sum_{\Lambda} a_{\Lambda} \mu_{\Lambda}^2$ with
the $\mu_{\Lambda}$'s satisfying $\sum_{\Lambda} \mu_{\Lambda}^2 =
1$. They can be parameterized by the angles on the 3-sphere as
explained in more detail in \cite{duff}. The remaining 4 angles
$\phi_{\Lambda}$ together with the $\mu_{\Lambda}$ describe the
internal space, while $x^{\mu}$ are coordinates of the
four-dimensional spacetime on which the resulting $N=2, D=4$
gauged supergravity is defined. The factors $a_{\Lambda}$ depend
on the four-dimensional axio-dilaton scalars $\tau_i =
e^{-\varphi_i} + i \chi_i$ (defined below) and the gauge fields
$A^{\Lambda} = A^{\Lambda}_{\mu} dx^{\mu}$ are exactly the ones
appearing in the four-dimensional theory. Note that if all the
gauge fields are vanishing and the scalars are at the minimum of
the potential, the internal space becomes exactly $S^7$. Apart
from the metric, the field strength of the 11-dimensional three
form field is given by:
\begin{align}\label{11dim_gauge}
\begin{split}
  F_{4} &= \sqrt{2} g \sum_{\Lambda} (a_{\Lambda}^2 \mu_{\Lambda}^2-\Delta
  a_{\Lambda})\epsilon_4+\frac{1}{\sqrt{2} g} \sum_{\Lambda}
  a_{\Lambda}^{-1} \bar{*} d a_{\Lambda} \wedge
  d(\mu_{\Lambda}^2)\\ &- \frac{1}{g^2} \sum_{\Lambda} a_{\Lambda}^{-2}
  d(\mu_{\Lambda}^2) \wedge (d\phi_{\Lambda}+\frac{g}{\sqrt{2}} A^{\Lambda}) \wedge \bar{*} d A^{\Lambda},
\end{split}
\end{align}
with $\bar{*}$ the Hodge dual with respect to the four-dimensional
metric $ds_4$, and $\epsilon_4$ the corresponding volume form.

With these identifications, the four-dimensional $N=2$ bosonic
Lagrangian, written in our conventions, reads
\begin{align}\label{duff_lagr}
\begin{split}
\mathcal L&=\frac{1}{2}R(g)+\frac{1}{4} \sum_{i = 1}^3
\left((\partial \varphi_i)^2+e^{2 \varphi_i} (\partial \chi_i)^2
\right) +
Im(\mathcal{M})_{\Lambda\Sigma}F_{\mu\nu}^{\Lambda}F^{\Sigma\,\mu\nu}
\\&+\frac{1}{2}Re(\mathcal{M})_{\Lambda\Sigma}\epsilon^{\mu\nu\rho\sigma}
F_{\mu\nu}^{\Lambda}F^{\Sigma}_{\rho\sigma} + 2 g^2 \sum_{i=1}^3
\left( \cosh \varphi_i + \frac{1}{2} \chi_i^2 e^{\varphi_i}
\right)\ .
\end{split}
\end{align}
One can then check explicitly (using also the particular result
for the matrix $\mathcal{M}$ given in \cite{duff}) that the above
Lagrangian is indeed of the form of \eqref{lagr} with prepotential
\begin{equation}\label{duff_prepot}
  F = - 2 i \sqrt{X^0 X^1 X^2 X^3}\ ,
\end{equation}
where the sections $X^{\Lambda}$ define the three scalars $\tau_i$
by $\frac{X^1}{X^0} \equiv \tau_2 \tau_3, \frac{X^2}{X^0} \equiv
\tau_1 \tau_3, \frac{X^3}{X^0} \equiv \tau_1 \tau_2$. The FI
parameters take the particularly simple form \begin{equation}
\xi_0 = \xi_1 = \xi_2 = \xi_3 = 1\ .
\end{equation} In this theory
one can find a black hole solution in analogy to the example in
section \ref{sect:5.1}. Following the general results in section
\ref{sect:4}, $X^{\Lambda} =
\alpha^{\Lambda}+\frac{\beta^{\Lambda}}{r}$, and from
\eqref{final_eq1}-\eqref{final_eq2} one can find the full solution
with $\alpha^0 = \alpha^1 = \alpha^2 = \alpha^3 = \pm\frac{1}{4}$
and three arbitrary parameters $\beta^1,\beta^2,\beta^3$ (or
equivalently $p^1,p^2,p^3$). We will not write down the full
solution as the expressions for the constant $c$ and the magnetic
charges in terms of the $\beta^{\Lambda}$'s are very long and do
not lead to further insight. It is clear that the particular
solution when we choose $\beta^1=\beta^2=\beta^3$ in fact
coincides precisely with the solution in section \ref{sect:5.1}
and this means that in any case a genuine black hole of the
M-theory reduction exists particularly when the three complex
scalars are equal. In the full solution of course there is a wider
range of values for $\beta^1,\beta^2,\beta^3$ that will lead to a
black hole, but this will suffice for our purposes here.

We now comment on the meaning of these four-dimensional black
holes from the point of view of M-theory as a first step towards
constructing the corresponding microscopic theory. It is notable
that the particular M-theory reduction we have leads to an
electrically gauged $N=2$ supergravity and thus the resulting
solution has only magnetic charges. This in fact makes the higher
dimensional interpretation a bit more involved. There are two main
points one can notice about the full 11-dimensional geometry from
the form in \eqref{11dim metric}. First, due to the nonconstant
scalars $\tau_i$, the full space is a warped product of the
internal seven-dimensional space with the AdS$_4$ black hole
spacetime. Second, due to the non-vanishing gauge fields
$A^{\Lambda}_{\varphi} = -p^{\Lambda} \cos \theta$, there is an
explicit mixing between the four angles $\phi_{\Lambda}$ of the
internal space and the four-dimensional angle $\varphi$. This
leads to four topological charges of the 11-dimensional spacetime,
in analogy to NUT charges. Note that in case the charges were only
electric, i.e.~$A^{\Lambda}_t = \frac{q^{\Lambda}}{r}$, the time
coordinate would mix with the internal angles and we would obtain
four angular momenta, leading to the interpretation of the
spacetime as arising from the decoupling limit of rotating
M2-branes as explained in detail in \cite{duff}. In the present
case however the interpretation of the four-dimensional black
holes from M-theory is more involved because apart from M2-branes
we need to have some Kaluza-Klein monopoles in the M-theory
solution, in order to account for the topological charge coming
from the magnetic charges in four dimensions. Unfortunately we
were not able to find an explicit example for this type of
solutions in the literature, which probably is also related to the
fact that they would break almost all supersymmetry\footnote{The
black hole solutions in four dimensions preserve only two
supercharges, i.e. they are 1/4 BPS in $N=2$. In $N=8$, they are
1/16 BPS. This means that at least 30 of the original 32
supercharges in the original 11-dimensional supergravity will have
to be broken for the conjectured bound state of M2-branes and
Kaluza-Klein monopoles.}.

\section{Black hole mass}\label{sect:Thermo}

In eqn.\eqref{proposedmass} we proposed a formula for the black
hole mass. In this section we provide more evidence for this using
using holographic renormalization. The computation is in fact
somewhat complicated due to the fact that it is hard to define an
energy, respectively mass, for asymptotically AdS black holes with
running scalars. A more detailed discussion on the complications
due to the scalars can be found in \cite{Barnich}. The correct
approach to the problem was developed in a series of papers
\cite{Kostas}, combining holographic regularization close to the
AdS boundary with the Hamilton-Jacobi method for finding the
appropriate counterterms. These results were collected by
\cite{Batrachenko} in a form we can readily use for our purposes.
For the particular class of black holes given by
\eqref{line-element_sol}, we can apply the formulas of
\cite{Batrachenko} and find the regularized energy to be
\begin{equation}\label{E_reg}
  E_{reg} = - 2 \omega_2 \left( g r_0 + \frac{c}{2 g r_0} \right)^2
  r_0 \left(-\frac{r_0}{2} \mathcal{K}'(r_0) + 1  \right),
\end{equation}
where $\omega_2$ is the volume element of a unit $2$-sphere and
the cutoff $r_0$ has to be eventually taken to infinity. This
expression clearly diverges, so one has to add to it the
counterterm energy given by
\begin{equation}\label{E_ct}
  E_{ct} = \omega_2 e^{-\mathcal{K}/2} \left( g r_0 + \frac{c}{2 g r_0} \right)
  \left(r_0^2 W(\phi)+ \frac{e^{\mathcal{K}}}{g} + \mathcal{O}(r_0^{-2}) \right).
\end{equation}
The expression $W(\phi)$ requires some further explanation. It
specifies the counterterms coming from the scalar fields and is
referred to as "superpotential" due to its resemblance with the
usual meaning of superpotential in supergravity. It should be
derived from the scalar potential via
\begin{equation}\label{superpotential}
  V = 2 G^{ij}(\phi) \frac{\partial W}{\partial \phi^i} \frac{\partial W}{\partial \phi^j} - \frac{3}{2}
  W^2.
\end{equation}
However, this expression does not rely on any supersymmetry and is
not necessarily unique as explained in more detail in
\cite{deBoer} in the five-dimensional case. The important point is
that one needs a set of real scalar fields $\phi^i$ which is not a
priori the case in $N=2$ supergravity. However, it might turn out
in practice that the solution effectively truncates the real or
imaginary part of the original complex scalars and thus one should
be in principle able to find the superpotential. This is indeed
what happens e.g. in the black hole solution coming from the $N=8$
truncation described above. Due to the importance of this
particular M-theory reduction, the theory was investigated and the
corresponding superpotential already found in \cite{Batrachenko}.
Let us first properly give the full solution in our conventions in
order to be able to describe precisely the relation between mass
and charges. We choose $X^1=X^2=X^3=\alpha+\frac{\beta}{r}$ as
explained in the previous section, and additionally have $X^0 =
\alpha^0 - \frac{3 \beta}{r}$. All the FI parameters are equal and
set to one, thus the BPS equations result eventually in the
following expression for the charges:
\begin{equation}
p^0 = \pm \frac{1}{g} \left(\frac{1}{4}-48 g^2 \beta^2 \right),
\qquad p^1 = p^2 = p^3 = \mp \frac{1}{g} \left(\frac{1}{4}-16 g^2
\beta^2 \right)\ ,\end{equation} for spinor I and II respectively.
The other constants in the solution are
\begin{equation}\alpha^0 = \pm \frac{1}{4}, \qquad \alpha = \pm \frac{1}{4}, \qquad c = 1 - 96 g^2 \beta^2\ .\end{equation}
The horizon is then found at $r_h = \sqrt{48 \beta^2-\frac{1}{2
g^2}}$ and requiring a genuine black hole with horizon constrains
$g \beta \in \left(-\infty, -\frac{1}{8}\right)$ for spinor I and
$g \beta \in \left(\frac{1}{8},+ \infty \right)$ for spinor II. Again, we find that
$c<-\frac{1}{2}$.
One can compute the superpotential to be
\begin{equation}W = \frac{g}{2} \left( \frac{(\pm r+ 4 \beta)^{3/4}}{(\pm r - 12 \beta)^{3/4}} + 3 \frac{(\pm r- 12 \beta)^{1/4}}{(\pm r + 4 \beta)^{1/4}} \right)\ .\end{equation}
Plugging this in \eqref{E_reg}-\eqref{E_ct} leads to
\begin{equation}E_{ren} = (E_{reg}+E_{ct})_{r_0 \rightarrow \infty} = \mp \omega_2 (512 g^2 \beta^3)\ ,\end{equation}
so we can define the mass to be
\begin{equation}\label{final_mass}
M = \mp 512 g^2 \beta^3\ ,
\end{equation}
which
is strictly bigger than zero for the black holes with horizon. In
fact we obtain the following interesting relation after plugging
in the possible ranges of $\beta$:
\begin{equation}
M > \frac{1}{g}\ . \end{equation} This inequality seems to be
generic enough independent of the technical details of the
particular solution, so we expect that it holds in general for the
new class of black holes.

It is very interesting to observe that the same value for the
black hole mass can be derived in a straightforward way from the
asymptotic expansion of the metric. In this case,
\begin{equation}
    U^2 (r\rightarrow\infty) = 4 g^2 r^2 + 4 (1-48 g^2 \beta^2) \pm \frac{1024 g^2
    \beta^3}{r}+ \mathcal{O} \left(\frac{1}{r^2} \right)\ .
\end{equation}
Assuming that the coefficient in front of $1/r$ is indeed $-2 M$
as in \eqref{RN-AdS}, we get back the same expression for the
mass, \eqref{final_mass}. This is a nice independent confirmation
that the procedure of holographic renormalization is well-defined.

As already shown in \cite{Kostas,Batrachenko}, the standard laws
of black hole thermodynamics hold with the above definition of
holographically renormalized energy. We can then summarize that
the new black hole solutions behave quite differently than the
usual case. All physical parameters of the solution are fixed in
terms of the gauge coupling $g$ and the constants $\beta$ (that
can be related to the charges). The solutions are singular in the
limits $g \rightarrow 0$ and $g \rightarrow \infty$. The limit of
large charges corresponds to large $\beta$ and this will be the
parameter that is easier to work with. Schematically, the physical
parameters in the large $\beta$ limit for fixed $g$ (i.e. fixed
AdS$_4$ radius) are
\begin{equation}(r_h - r_s) \sim \beta, \qquad p \sim \beta^2, \qquad M \sim \beta^3, \qquad S \sim \beta^2\ ,\end{equation}
with $r_h - r_s$ the radius of the black hole. It is then clear that the entropy in fact scales linearly with
the charges, while the mass scales as $p^{3/2}$. This behavior is very atypical for black holes and it would be
interesting to justify it on more general grounds from the supersymmetry algebra in AdS.

\section{Outlook}

From this work it should be clear that one implicit assumption about solutions in gauged supergravities is in
fact incorrect. There do exist qualitatively very different types of spacetime solutions in gauged supergravity
with vector multiplets compared to the minimally gauged supergravity case. As examples, we discussed
supersymmetric, static AdS$_4$ black holes with spherical symmetry in gauged supergravity with Fayet-Iliopoulos
terms. To achieve a full classification of black holes in gauged supergravity, one has to consider a general
supergravity setup with arbitrary number of vector multiplets and also potentially hypermultiplets and tensor
multiplets. The present work is in this respect a small step towards a broader understanding of such black hole
solutions.

Further it is clear that the solutions described here are a very
particular type and one can imagine different extensions to, e.g.,
rotating $1/2$ and $1/4$ BPS black holes with nontrivial scalars
along with higher dimensional analogues of the static solution.
The role and exact meaning of the attractor mechanism in AdS$_4$
black holes must also be better understood, an issue related with
the construction of M-brane or D-brane description of these black
holes. In this sense it is important to understand clearly the
physical reason why the entropy of the black holes depends also on
the gravitino charges. The thermodynamic description and the
precise BPS bounds also need a more solid basis. It will be
interesting to extend the present solutions also to extremal
non-BPS and finite temperature analogues. We hope to address at
least some of these issues in the future.

\section*{Acknowledgments}
We would like to thank G. Barnich, B. de Wit, S. Katmadas, D.
Klemm, P. Nedkova, C. Toldo, and T. Ort\'{i}n for helpful
discussions and correspondence. We acknowledge support by the
Netherlands Organization for Scientific Research (NWO) under the
VICI grant 680-47-603.
\appendix

\section{Gamma matrix conventions}\label{app:A}
The Dirac gamma-matrices satisfy
\begin{align}
\begin{split}
  \{\gamma_a,\gamma_b\} &= 2 \eta_{ab}\ ,\\
  [\gamma_a,\gamma_b] &\equiv 2 \gamma_{ab}\ ,\\
\gamma_5 &\equiv - i \gamma_0 \gamma_1 \gamma_2 \gamma_3 = i
\gamma^0\gamma^1 \gamma^2 \gamma^3\ .
\end{split}
\end{align}
In addition, they can be chosen such that
\begin{align}
  \gamma_0^\dagger = \gamma_0, \quad \gamma_0 \gamma_i^\dagger
  \gamma_0 = \gamma_i,\quad \gamma_5^\dagger = \gamma_5,\quad   \gamma_\mu^* = -\gamma_\mu\ .
\end{align}
An explicit realization of such gamma matrices is the Majorana
basis, given by
\begin{align}
  \gamma^0 &= \begin{pmatrix}0 & \sigma^2\\ \sigma^2 &
    0 \end{pmatrix},&
  \gamma^1 &= \begin{pmatrix}i \sigma^3 & 0\\ 0&i\sigma^3 \end{pmatrix},&
  \gamma^2 &= \begin{pmatrix}0 & -\sigma^2\\ \sigma^2 &
    0 \end{pmatrix},\nonumber\\
  \gamma^3 &= \begin{pmatrix}-i \sigma^1 & 0\\
    0&-i\sigma^1 \end{pmatrix},&
  \gamma_5 &= \begin{pmatrix}\sigma^2 & 0\\ 0&-\sigma^2 \end{pmatrix}\ ,
\end{align}
where the $\sigma^i; i=1,2,3$ are the Pauli matrices. Their
$SU(2)$ matrix indices $A,B$ can be lowered or raised with the
antisymmetric tensor. We then obtain the following set of
matrices:
\begin{equation}
\sigma^1 = \left(
\begin{array}{cc}
0 & 1  \\
1 & 0  \end{array} \right), \quad \sigma^2 = \left(
\begin{array}{cc}
0 & -i  \\
i & 0  \end{array} \right), \quad \sigma^3 = \left(
\begin{array}{cc}
1 & 0  \\
0 & -1  \end{array} \right), \quad indices_A{}^B.
\end{equation}

\begin{equation}
\sigma^1 = \left(
\begin{array}{cc}
1 & 0  \\
0 & -1  \end{array} \right), \quad \sigma^2 = \left(
\begin{array}{cc}
-i & 0  \\
0 & -i  \end{array} \right), \quad \sigma^3 = \left(
\begin{array}{cc}
0 & -1  \\
-1 & 0  \end{array} \right), \quad \epsilon = \left(
\begin{array}{cc}
0 & 1  \\
-1 & 0  \end{array} \right), \quad indices_{A B}.
\end{equation}

\begin{equation}
\sigma^1 = \left(
\begin{array}{cc}
-1 & 0  \\
0 & 1  \end{array} \right), \quad \sigma^2 = \left(
\begin{array}{cc}
-i & 0  \\
0 & -i  \end{array} \right), \quad \sigma^3 = \left(
\begin{array}{cc}
0 & 1  \\
1 & 0  \end{array} \right), \quad \epsilon = \left(
\begin{array}{cc}
0 & 1  \\
-1 & 0  \end{array} \right), \quad indices^{A B}.
\end{equation}
Notice the property $(\sigma^i_{AB})^*=-\sigma^{i\,AB}$.

\end{document}